# Structure of Turbulence in Katabatic Flows below and above the Wind-Speed Maximum

**Andrey A. Grachev • Laura S. Leo • Silvana Di Sabatino • Harindra J. S. Fernando • Eric R. Pardyjak • Christopher W. Fairall**

---

Andrey A. Grachev (✉)
NOAA Earth System Research Laboratory / Cooperative Institute for Research in Environmental Sciences, University of Colorado, 325 Broadway, R/PSD3, Boulder, CO 80305-3337, USA
e-mail: Andrey.Grachev@noaa.gov

Andrey A. Grachev • Laura S. Leo • Silvana Di Sabatino • Harindra J. S. Fernando
Department of Civil & Environmental Engineering & Earth Sciences, University of Notre Dame, Notre Dame, IN, USA

Silvana Di Sabatino
Department of Physics and Astronomy, University of Bologna, Bologna, Italy

Eric R. Pardyjak
Department of Mechanical Engineering, University of Utah, Salt Lake City, UT, USA

Christopher W. Fairall
NOAA Earth System Research Laboratory, Boulder, CO, USA

**Abstract**

Measurements of small-scale turbulence made over the complex-terrain atmospheric boundary layer during the Mountain Terrain Atmospheric Modeling and Observations (MATERHORN) Program are used to describe the structure of turbulence in katabatic flows. Turbulent and mean meteorological data were continuously measured at multiple levels (up to seven) at four towers deployed along the East lower slope (2-4 degrees) of Granite Mountain. The multi-level observations made during a 30-day long MATERHORN–Fall field campaign in September-October 2102 allowed studying of temporal and spatial structure of katabatic flows in detail, and herein we report turbulence statistics (e.g., fluxes, variances, spectra, and cospectra) and their variations in katabatic winds. Observed vertical profiles show steep gradients near the surface, but in the layer above the slope jet the vertical variability is smaller. It is found that the vertical (normal to the slope) momentum flux and horizontal (along the slope) heat flux in a slope-following coordinate system change their sign below and above the wind maximum of a katabatic flow. The vertical momentum flux is directed downward (upward) whereas the horizontal heat flux is downslope (upslope) below (above) the wind maximum. Our study therefore suggests that the position of the jet-speed maximum can be obtained by linear interpolation between positive and negative values of the momentum flux (or the horizontal heat flux) to derive the height where flux becomes zero. It is shown that the standard deviations of all wind speed components (therefore the turbulent kinetic energy) and the dissipation rate of turbulent kinetic energy have a local minimum, whereas the standard deviation of air temperature has an absolute maximum at the height of wind-speed maximum. We report several cases where the vertical and horizontal heat (buoyancy) fluxes are compensated. Turbulence above the wind-speed maximum is decoupled from the surface, and follows the classical local $z$-less predictions for stably stratified boundary layer.

# 1 Introduction

The local circulation in mountainous areas in part can be generated by vertical density gradients on sloped terrains (e.g. Whiteman 2000). During the night time in mid-latitudes or at high latitudes the atmospheric boundary layer is often stably stratified, and on sloping terrain downslope flows (or katabatic flows) are generated above the surface. Katabatic flows are common over glaciers and ice sheets in Antarctica or Greenland.

A prominent feature of katabatic flow is a wind maximum close to the surface that causes a sign change in the momentum flux below and above the wind maximum. This, obviously, limits the application of traditional approaches for flux-profile relationships derived for the stable boundary layer (SBL) over flat surfaces where the vertical gradient of mean wind speed is considered to have the same sign. The downslope low-level jet is triggered by the positive vertical density gradient on a sloping surface which also acts horizontally (along the slope) as katabatic forcing. Generally, the katabatic forcing term in the momentum budget equation is smaller than other terms (e.g. background horizontal pressure gradient) and, for this reason, katabatic flows are generally observed during quiescent periods in the SBL. Under such conditions, katabatic flows efficiently drive the turbulent exchange of momentum, heat, moisture, and pollutants between the earth's surface and the atmosphere. However, katabatic flows are poorly resolved in most numerical weather prediction, climate, and air pollution models because the typical jet maximum is located close to the earth's surface.

Though much work has already been carried out on katabatic flows, a unified picture or theory does not exist. Several analytical models have been proposed, one of the first being the classical analytical solutions of Prandtl (1942), a case that Mahrt (1982) calls 'equilibrium flows'. Prandt's approach has been extended to include time dependence, Coriolis effects, height-dependent eddy viscosity and diffusivity coefficients etc. (e.g. Lykosov and Gutman 1972; Egger 1990; Ingel' 2000; Grisogono and Oerlemans 2001a,b; Grisogono 2003; Parmhed et al. 2004; Kavčič and Grisogono 2007; Stiperski et al 2007; Shapiro and Fedorovich 2008; and references therein). The Prandtl model provides a conceptual picture of katabatic flows and, when appropriately tuned, can qualitatively reproduce the observed vertical profiles of wind speed and potential temperature (e.g. Papadopoulos et al. 1997; Grisogono and Oerlemans 2001a). This model, however, is a simple analytical one-dimensional framework and it has glaring deficiencies (see Mahrt 1982; Oerlemans and Grisogono 2002; Zammett and Fowler 2007; Axelsen et al. 2010; Grisogono and Axelsen 2012, for discussion). This has prompted the use of numerical models such as closure models, large-eddy simulation (LES), and direct numerical simulation (DNS), which are based on the complex governing equations that include the non-linear terms. Numerical weather prediction models have also been widely used to study katabatic flows (e.g. Renfrew 2004 and references therein).

In several papers 1.5-order and second-order closure models for the turbulent kinetic energy (TKE) have been used to study katabatic winds (see Denby 1999, Table I for survey). Whereas Denby (1999) successfully simulated both mean and flux profiles of a katabatic flow using second-order closure, such approaches require physical assumptions and frequently tuning of empirically determined model constants, resulting in a large uncertainty and limited general applicability.

Direct numerical simulation (DNS) of turbulent katabatic flows with and without the Coriolis effect was conducted by Shapiro and Fedorovich (2008) and Fedorovich and Shapiro (2009). In DNS, no parameterization is used, but they are largely confined to idealized flows and hence generally cannot resolve the largest turbulent eddies. This is because, in DNS, the governing equations of turbulence are solved numerically for low to moderately high Reynolds numbers (of the order $10^4$), but in the atmosphere the Reynolds numbers are several orders of magnitude larger than in DNS.

LES was introduced as an intermediate solution (between a closure model and DNS). In LES, the motions of the large eddies related to the production and transport of turbulence are explicitly resolved, while only the small-scale turbulent eddies require modelling (sub-grid scale turbulence) that brings an uncertainty in outcome of LES models. The development of LES models during the last decades has enabled simulation of boundary layer flows such as katabatic winds. Recently, Skyllingstad (2003), Smith and Skyllingstad (2005), Axelsen and van Dop (2009a, b), Grisogono and Axelsen (2012) used LES to simulate katabatic flows (see a review of different LES by Smith and Porté-Agel 2013, their Table 1).

Katabatic winds have a long history of investigation and the relevant literature is voluminous. Katabatic winds have been experimentally described covering various regions of the world, including European Alps (e.g., Nadeau et al. 2013a, b; Oldroyd et al. 2014), Greece (Helmis and Papadopoulos 1996; Papadopoulos et al. 1997), Spain (Viana et al. 2010), the US Mountain States and Southwest (e.g., Horst and Doran 1986, 1988; Neff and King 1987, 1988; Clements et al. 1989; Stone and Hoard 1989; Monti et al. 2002, 2014; Haiden and Whiteman 2005; Princevac et al 2005, 2008; Whiteman and Zhong 2008; Pardyjak et al. 2009), Australia (Manins and Sawford 1979), over different glaciers and Polar ice caps and sheets (e.g., Meesters et al. 1997; Van den Broeke 1997; Smeets et al. 1998, 2000; Oerlemans et al. 1999; Van der Avoird and Duynkerke 1999; Denby and Smeets 2000; Renfrew and Anderson 2006) etc.

Detailed review of observational history of katabatic winds can be found in Poulos and Zhong (2008).

Limited observations still remain a problem for validation and calibration of katabatic flow models. In particular, during past field campaigns, turbulent measurements of the katabatic flows were generally limited by a single flux tower equipped with a few levels (and rarely several levels) of sonic anemometers. These conditions made description of the turbulence structure of katabatic flows difficult. The turbulence data collected in mountain terrain during the M̲ou̲nt̲ain T̲er̲rain Atmosp̲h̲eric M̲o̲deling and Obs̲e̲rvatio̲n̲s (MATERHORN) campaign offer several advantages for studying the katabatic flows compared to previous field programs. These long-term, multi-level, multi-tower turbulent observations of the nocturnal SBL allow us to study the turbulence structure of the katabatic flows in detail. Here we report some results of turbulent measurements from the first MATERHORN field campaign (MATERHORN–Fall) carried out at the US Army Dugway Proving Grounds in Utah from 25 September through 31 October, 2012 (Fernando et al. 2014).

**2 The TKE Equation in a Slope-Following Coordinate System**

In this section we briefly describe the turbulent kinetic energy (TKE) equation. Unlike boundary layers over flat horizontal surfaces, the governing equations of the katabatic winds are described in a Cartesian coordinate system aligned with the slope, which is inclined at an angle $\alpha > 0$ to the horizontal (e.g., Denby 1999; Shapiro and Fedorovich 2008, 2014; Axelsen and Van Dop 2009a; Fedorovich and Shapiro 2009; Grisogono and Axelsen 2012; Łobocki 2014). The transformation from the traditional coordinate system where the vertical axis is aligned with the force of gravity to a slope-following coordinate system (i.e., rotation of the reference frame around the cross-slope axis by the slope angle $\alpha$) can be accomplished by use of the metric tensor and the vector of the gravity field applied to the original equations (see details in Denby 1999, Eq. (18); Łobocki 2014, Appendix 2). In the current study, the katabatic flows are considered in a slope-following right-hand Cartesian coordinate system with axes directed, respectively, down the slope, across the slope, and perpendicular to the slope. Hereinafter, normal to a slope and along a slope fluxes are referred to as vertical and horizontal fluxes respectively (if it is not stipulated specifically).

In a rotated coordinate system, the governing equations contain several modifications. In particular, the equation for downslope momentum contains the so-called katabatic forcing term associated with the temperature (density) perturbations (e.g. Mahrt 1982). As mentioned above, this term by definition drives the katabatic flow. In some sense, the katabatic forcing term is similar to the net pulling force in the classical physics for the inclined plane (ramp) problem (this force is causing a block to slide down along the slope). Another important modification is associated with the TKE equation which we consider in more detail because our study focuses on observation of small-scale turbulence. In a rotated streamline coordinate system, the TKE equation becomes (e.g., Horst and Doran 1988; Denby 1999; Łobocki 2014):

$$\partial <e>/\partial t = -<u'w'>(\partial U/\partial n) + \beta(<w'\theta_v'>\cos\alpha - <u'\theta_v'>\sin\alpha) - T - \varepsilon . \quad (1)$$

Here $e = (u'^2 + v'^2 + w'^2)/2$ is the TKE, $U$ is mean along-slope wind speed, $n$ is the coordinate normal to the slope, $\varepsilon$ is the dissipation rate of the TKE, $\theta_v$ is the virtual potential temperature, $\beta = g/\theta$ is the buoyancy parameter ($g$ is the acceleration due to gravity and $\theta$ is the air temperature), $u$, $v$, and $w$ are the longitudinal (down-slope), lateral (cross-slope), and vertical (normal) velocity components, respectively, [′] denotes fluctuations about the mean value, and $< >$ is a time or space averaging operator. The transport and pressure work term in (1) is defined by $T = \partial(<w'e> + <w'p'>/\rho)/\partial n$ where $p'$ is the fluctuation in atmospheric pressure and $\rho$ is the air density.

In a slope-following coordinate system, the buoyancy term in the TKE equation (1) is changed. An additional part $\beta <u'\theta'_v> \sin\alpha$ is associated with the rotation around the gravity vector and the "horizontal" (along the slope) buoyancy flux (or sensible heat flux in the case of the dry air). The modification of the buoyancy term in Eq. (1) leads to a modification of several stability parameters which contain this term. In a slope-following coordinate system, the flux Richardson number is (cf. Łobocki 2014, Eq. (23))

$$Rf = -\frac{\beta(<w'\theta'_v>\cos\alpha - <u'\theta'_v>\sin\alpha)}{<u'w'>(\partial U/\partial n)} \ . \qquad (2)$$

The Monin-Obukhov stability parameter in a rotated coordinate system is defined as the ratio of a reference height $n$ normal to the slope and a modified Obukhov length scale $L$:

$$\zeta \equiv \frac{n}{L} = -\frac{n\,\kappa\,\beta(<w'\theta'_v>\cos\alpha - <u'\theta'_v>\sin\alpha)}{u_*^3} \ . \qquad (3)$$

where the friction velocity $u_* = (<u'w'>^2 + <v'w'>^2)^{1/4}$ is considered positive below and above the wind maximum of slope flow. The von Kármán constant $\kappa \approx 0.4$ is included in Eq. 3 simply by convention. Discussion on importance of the $\beta <u'\theta'_v> \sin\alpha$ term to the Monin-Obukhov stability parameter (3) can be also found in Horst and Doran (1988, p. 615). Note that the sign of the $\sin\alpha$ factor in the buoyancy term in (1)–(3) depends on the direction of the along-slope axis (e.g., Shapiro and Fedorovich 2014, their Footnote 3). The sign is negative if the along-slope axis points down the slope (Horst and Doran 1988 and the current study) and vice versa (Denby 1999; Łobocki 2014 papers).

The additional term $\beta <u'\theta'_v> \sin\alpha$ in the TKE budget can enhance or suppress turbulence (depending on its sign which will be discussed shortly), leading to a change in the critical gradient and flux Richardson numbers, which may not coincide with the canonical 'critical value' of 0.20 or 0.25 obtained for flat horizontal surface (see Grachev et al. 2013 for discussion). The critical value of the gradient and flux Richardson numbers for the katabatic flows depends on a slope angle and the TKE budget (Horst and Doran 1988; Denby 1999). Near a local wind-speed maximum, the shear term becomes small and the gradient Richardson number can reach very high values, up to $Ri = 200$ (Smeets et al. 2000, their Fig.4; Tse et al. 2003, their Fig. 3; Söderberg and Parmhed 2006, their Fig. 10).

Existence of a wind maximum in katabatic flows leads to a sign reversal of the vertical momentum flux and horizontal heat flux at the wind-maximum height. For stably stratified flow over sloping terrain, the vertical gradient of mean potential temperature is positive in the entire layer, $d\theta/dn > 0$ (in the general case $d\theta_v/dn > 0$). However, the vertical gradient of mean wind speed is positive, $dU/dn > 0$, below the wind maximum and it is negative, $dU/dn < 0$, above the wind maximum (obviously that $dU/dn = 0$ at the wind-maximum height). To understand vertical behaviour of the turbulent moments, we use physical arguments based on an idealized turbulent eddy mixing. In the layer below the wind maximum, an upward moving air parcel ($w' > 0$) ends up being slower ($u' < 0$) and cooler ($\theta' < 0$) than its surroundings, while a downward ($w' < 0$) moving air parcel is faster ($u' > 0$) and warmer ($\theta' < 0$). We are assuming that particle temperature and velocity are conserved during its travel. Thus in this region, both the upward and downward moving air parcels contribute negatively to the both $w\theta$- and $uw$- covariances, that is, $<w'\theta'> < 0$ and $<u'w'> < 0$ respectively (meaning a downward transport of heat and momentum), but contribute positively in the horizontal heat flux, $<u'\theta'> > 0$.

However, the net fluid transport is zero ($<w> = 0$) as expected from the equation of continuity. In the region above the slope flow wind maximum ($dU/dn < 0$ and $d\theta/dn > 0$), an upward moving air parcel ($w' > 0$) is faster ($u' > 0$) and cooler ($\theta' < 0$) than its surroundings, while a downward ($w' < 0$) moving parcel is slower ($u' < 0$) and warmer ($\theta' < 0$), resulting in $<w'\theta'> < 0$, $<u'w'> > 0$, and $<u'\theta'> < 0$ for both the upward and downward moving air.

Thus, the $<u'\theta'_v>$ term is positive or downslope (a sink for TKE) below the wind-speed maximum and negative or upslope (a source for TKE) above. Because the vertical flux $<w'\theta'_v>$ term is always negative under the stable conditions (a sink for TKE), it is therefore the horizontal heat flux $<u'\theta'_v>$ term increases (decreases) stability parameters (2) and (3) below (above) the wind-speed maximum. The contribution of the $<w'\theta'_v>\cos\alpha$ term to the modified buoyancy term in the TKE equation (1) decreases with slope angle, while the contribution of the $<u'\theta'_v>\sin\alpha$ term increases with slope angle. In the case where

$$-<u'\theta'_v> \;>\; -\cot\alpha <w'\theta'_v>, \tag{4}$$

the net buoyancy term in Eq. 1 will be always positive even if the surrounding flow is stably stratified, that is, the inequality (4) implies a net positive buoyancy production of the TKE (recall that the both vertical and horizontal heat fluxes are negative). Note that condition (4) is possible only in the region above the wind maximum and it is never reached under the wind maximum (see also field data presented in Sect. 4.1). The limit $<u'\theta'_v>/<w'\theta'_v> = \cot\alpha$ implies that the total heat flux vector is perpendicular to the gravity vector (Denby 1999). In other words a sink of TKE due to vertical buoyancy is completely cancelled by a source of TKE due to the tilted horizontal buoyancy. Historically the inequality (4) was suggested and discussed by Denby (1999). According to Denby (1999, p. 79), Eq. 4 results in the critical angle $\alpha \approx 25°$ for $<u'\theta'_v> / <w'\theta'_v> \approx 2.1$. A ratio of horizontal to vertical heat fluxes derived from the MATERHORN–Fall data will be discussed further in the Section 4. Discussion on importance of the $<u'\theta'_v>$ term in the other second-moment equations can be found in Horst and Doran (1988, p. 613).

Although in the layer above the flow wind maximum, the kinematic momentum flux is negative, $\tau = -<u'w'> < 0$ (upward momentum transfer), the production of turbulence by the mean flow shear in the TKE budget equation (1), $-<u'w'>(\partial U / \partial n)$, and the turbulent viscosity, $K_m = -\dfrac{<u'w'>}{dU/dn}$, are positive because both the momentum flux and gradient of mean wind speed change sign simultaneously. The change in the sign of the momentum flux for the slope flows was theoretically and experimentally reported and discussed by Horst and Doran (1988), Neff (1990), Denby (1999), Denby and Smeets (2000), Söderberg and Tjernström (2004), Kouznetsov et al. (2013), Nadeau et al (2013b), Monti et al. (2014), and Oldroyd et al. (2014).

Note that the phenomenon of the upward momentum transfer is well known in air-sea interaction (e.g. Grachev and Fairall 2001 and references therein). Upward momentum transfer in the marine boundary layer (i.e., from water to air) is usually associated with fast-traveling swell running in the same direction as the wind or with decaying wind conditions (e.g. after the passage of a storm or gale) when the wind speed is less than the phase speed of the wave spectral peak. This regime can be treated as swell regime or mature sea state. Such fast waves lead to a low-level wave-driven wind jet (Harris 1966). Based on simultaneous airborne and mast observations, Smedman et al. (1994) found that upward momentum flux can occupy the lowest 200 m of the atmospheric boundary layer. Hanley and Belcher (2008) reported that in the layer with the upward moment transfer, wind turns in the opposite direction to the classical Ekman spiral. A low-level wave-driven wind jet (e.g. Hanley and Belcher 2008, their Fig. 9) and a coastal jet (Brooks et al. 2003, their Figs. 3b and 8a) are very similar to a slope low-level jet, and, one may expect, that study of katabatic flows can be useful for the problem of the swell regime over oceans or jet-like structures and vice versa.

Another issue is the transformation of coordinate system in terms of measurements, considering that all theoretical results are derived in a slope-following coordinate system; however, theoretical or model findings are generally compared with the experimental data obtained in non-rotated Cartesian coordinate system where a vertical axis is aligned with the force of gravity (the current study is not an exclusion). For example, vertical gradients of air temperature/humidity or mean wind speed (measured by cup anemometers) are derived from sensors aligned with the gravity vector ("true" vertical line). Although the mean wind speed, direction, and turbulent wind stress are derived from a sonic anemometer with double rotation of the anemometer axes needed to place the measured wind components in a terrain-following coordinate system (see details in the next section), the origin of these vectors in the case of multi-level measurements are also located on a "true" vertical line ("mixed" coordinate system).

One can assume that for katabatic flows over gentle terrain, the discrepancy between measurements in a slope-following coordinate system and in a non-rotated coordinate system is insignificant and is within the accuracy of the experimental data. However, this difference may be substantial over very steep (e.g. $\alpha = 20°$- $40°$) slopes (cf. Geissbuhler et al. 2000; Van Gorsel et al. 2003; Nadeau et al. 2013a, b; Oldroyd et al. 2014). We suggest that this intricacy should be taken into account in future field campaigns over steep slopes by modifying the experimental setup. For example, 'slow' T/RH probes, sonic anemometers, and other sensors can be aligned with a line normal to the slope using mounting arms/booms with different lengths (arms at upper levels should be longer than arms at lower levels) whereas a tower can still be aligned with the gravity vector. A length difference $\Delta l$ of two arms located at different measurement levels should be $\Delta l = \Delta z \tan \alpha$ (if the arms are aligned with the "true" horizontal direction, that is, perpendicular to the tower) and $\Delta l = \Delta z \sin \alpha$ (if the arms are aligned with the slope) where $\Delta z$ is height difference between the two levels (in a non-rotated coordinate system) and $\alpha$ is a slope angle. Another issue is the azimuth and angle-of-attack dependent errors due to sensor orientation relative to the flow when the arms and/or sonic anemometers over a slope are aligned with the "true" horizontal direction (see Geissbuhler et al. 2000; Van Gorsel et al. 2003; Kochendorfer et al. 2012; Mauder 2013; Nadeau et al. 2013a for discussion).

**3 The MATERHORN Observation Site and Instrumentation**

The MATERHORN program is a five-year effort designed to better understand flow and turbulence process in mountainous terrain for improved mesoscale modelling and weather predictability. This is an Office of Naval Research (ONR) funded Multidisciplinary University Research Initiative (MURI) of the Department of Defense (DoD) project led by the University of Notre Dame, Indiana. The program includes four major components: modelling, field experimental, technology development, and parameterization components. A comprehensive experimental component (MATERHORN–X) focuses on field measurements for studying atmospheric processes within complex-terrain. The plans called for two major campaigns with high resolution measurements. The campaign periods were selected based on the climatology of the area. The fall campaign (MATERHORN–Fall, September - October, 2012) focused on quiescent, dry, fair weather (wind speeds < 5 m/s) periods dominated by diurnal heating/cooling, and the spring campaign (MATERHORN–Spring, May, 2013) sought measurements under strong synoptic conditions. Both MATERHORN–X field campaigns were carried out at Granite Mountain Atmospheric Science Testbed (GMAST) of the Dugway Proving Grounds, a US Army facility, located approximately 85 miles (140 km) southwest of Salt Lake City, Utah in southern Tooele County and just north of Juab County. General information about the MATERHORN program and the field experiments can be found in Fernando et al. (2014).

Granite Mountain, an isolated topography within the DPG, is the centerpiece of the MATERHORN–X program. The length of Granite Mountain is 11.8 km; the largest width is 6.1 km and peak elevation 0.84 km above the valley elevation (1.3 km above the sea level). Granite Mountain was surrounded by several Intensive Observing Sites (IOS) including IOS-ES (East Slope) and IOS-WS (West Slope) to study slope flows, their interaction with valley flows, flow oscillations, and canyon effects. All IOSs had heavily instrumented towers, at least one 20 m in height. To examine katabatic flows in detail, five towers designated as ES1-ES5 (IOS-ES) were placed along the flow line on the east slope of Granite Mountain and separated by about 600-700 m (Fig. 1).

The present study uses the data collected at IOS-ES (towers ES2 to ES5 only) during the experiment MATERHORN–Fall in the fall of 2012. The towers ES2–ES5 were instrumented with fast response three-axis sonic anemometer/thermometers that sampled at 20 Hz and slow response temperature and relative humidity (T/RH) probes that sampled at 1 Hz on the ES2 and ES3 towers and at 0.5 Hz on the ES4 and ES5 towers. Each flux tower at IOS-ES had several (at least five) levels of measurements. The sonic anemometers and the 'slow' T/RH probes were placed at seven levels on the ES2 tower (0.5, 4, 10, 16, 20, 25, and 28 m), at five levels on the ES3 tower (0.5, 2, 5, 10, and 20 m), at six levels on the ES4 tower (0.5, 2, 5, 10, 20, and 28 m), and at five levels on the ES5 tower (0.5, 2, 5, 10, and 20 m). The ES2 and ES4 towers were instrumented entirely with R.M. Young (Model 81000) sonic anemometers whereas the ES5 tower was instrumented entirely with Campbell Scientific, Inc. CSAT3 sonic anemometers. The ES3 tower was instrumented with Campbell CSAT3 sonic anemometer and fast response Campbell KH20 krypton hygrometer at 2-m level and with R.M. Young sonic anemometers at other measurement levels. The towers were placed along the flow line on the east-facing slope of Granite Mountain (Fig. 1) with inclinations ranging approximately in the east-west direction from 2 to 4 degrees gradually increasing from ES2 to ES5 tower (Fig. 2).

The 'slow' T/RH probes provided air temperature and relative humidity measurements at several levels and were used to evaluate the vertical temperature and humidity gradients based on half-hour averaged 1-Hz data. The mean wind speed and wind direction were derived from the sonic anemometers, with rotation of the anemometer axes needed to place the measured wind components in a streamline coordinate system based on half-hour averaged 20-Hz data. We used the most common method, which is a double rotation of the anemometer coordinate system, to compute the longitudinal, lateral, and vertical velocity components in real time (Kaimal and Finnigan 1994, Sect. 6.6).

The 'fast' 20-Hz raw data measured by a sonic anemometer were first edited to remove spikes from the data stream. Turbulent covariance and variance values were derived through the frequency integration of the appropriate cospectra and spectra computed from 27.31-min data blocks (corresponding to $2^{15}$ data points) from the original 1/2 hour data files. In addition, to separate the contributions from mesoscale motions to the calculated eddy-correlation flux, a low-frequency cut-off at 0.0076 Hz (the tenth spectral value or a period of about 2 min) was applied on the cospectra as a lower limit of integration; the upper limit of integration is 10 Hz (the Nyquist frequency). The low-frequency cut-off for turbulent contributions is chosen to lie in the spectral gap between the small- and large-scale contributions to the total transport (see Grachev et al. 2013, 2014 for detail).

The dissipation rate of turbulent kinetic energy $\varepsilon$ in Eq. 1 was estimated based on a common method for measuring $\varepsilon$ in a turbulent flow that assumes the existence of an inertial subrange associated with a Richardson-Kolmogorov cascade. The frequency energy spectrum of the longitudinal velocity component, $S_u(f)$, in the inertial subrange has the form:

$$S_u(f) = \alpha_K (U/2\pi)^{2/3} \varepsilon^{2/3} f^{-5/3}, \qquad (5)$$

where $f$ is the frequency, $U$ is mean wind speed, and $\alpha_K$ is the Kolmogorov constant with a value estimated $\alpha_K \approx 0.5$-$0.6$ (e.g. Kaimal and Finnigan 1994); a value $\alpha_K = 0.55$ is adopted for the current study. If the turbulence is locally isotropic, the spectra of lateral and vertical velocity components are 4/3 of the longitudinal velocity; that is,

$$S_v(f) = S_w(f) = (4/3) S_u(f). \qquad (6)$$

Based on (5) and (6), we derived $\varepsilon$ separately from the spectra for each velocity component ($u'$, $v'$, and $w'$) in the frequency domain 0.9–2.7 Hz (between the 50th and 61st spectral values) located within the inertial subrange. The median of these three values is taken as the representative dissipation rate. With this procedure, the influence of possible spectral spikes on the estimation of the dissipation rate and reduced sampling error is averted (see Grachev et al. 2014 and references therein for discussion). Because our estimates of $\varepsilon$ are based on Eqs. (5) and (6), data without the Richardson-Kolmogorov cascade should be filtered out. In the current study, the following prerequisite is imposed on the data. The data points where the spectral slope in the inertial subrange (in the frequency domain 0.9–2.7 Hz) deviated more than 20% of the theoretical -5/3 slope were excluded from the analysis (cf. Hartogensis and De Bruin, 2005, where ±20% was also used).

Similarly, the dissipation (destruction) rate for half the temperature variance, $N_t$, was derived from –5/3 Obukhov-Corrsin power law for a passive scalar

$$S_t(f) = \beta_K (U/2\pi)^{2/3} N_t \varepsilon^{-1/3} f^{-5/3} , \qquad (7)$$

where $\beta_K$ is the Kolmogorov (Obukhov-Corrsin) constant for a passive scalar; a value $\beta_K = 0.8$ (e.g. Kaimal and Finnigan 1994) is used in the current study. The dissipation rate of turbulent kinetic energy $\varepsilon$ in Eq. 7 is estimated first from Eqs. 5 and 6 as described above.

**4 Observed Turbulent Structure of Katabatic Flows**

This paper concerns only those flows that resemble "pure" katabatic flows simultaneously at all ES2-ES4 flux towers, meaning that all profiles have a low-level wind maximum. Our observations during the MATERHORN–Fall show that the katabatic flows are associated with quiescent synoptic conditions and generally clear skies. These flows are remarkably unidirectional and its duration can reach 2-3 hours. It is found that the katabatic flows on the East slope of Granite Mountain ("slope flows") are rather intermittent and often disturbed due to strong interaction between the slope flows and the circulation in the Dugway valley occurring at various times during the night (Hocut et al. 2014). The westerly slope flows develop rapidly soon after sunset when the surface starts to cool; they persist for more than two hours, interrupt, arise again, and decay at dawn.

Although episodes of the katabatic winds over the East slope occur quite often, only six cases of persistent westerly katabatic winds observed during the three night of 28, 30 September and 2 October 2012 (YD = 272, 274, and 276) are analyzed in the current study. During these nights, the episodes of katabatic winds are observed at all ES2-ES4 flux towers from about 0200 to 0400 UTC and from 0530 to 0630 UTC (local time is from 2000 to 2200 and from 2330 to 0030 of a previous day for the most part). Note that the local time in Utah during the experiment MATERHORN-Fall is UTC/GMT minus 6 hours, that is, time zone is the US Mountain Daylight Time (MDT). All times hereinafter are time stamped to reflect a 1/2 hour data file. For example, a date time 0200 indicates that data were collected and averaged from 0200 until 0230.

One may suggest that the identical time periods of the observed katabatic winds on the East slope of Granite Mountain during these nights may characterize universal pattern of nocturnal circulation at the Dugway basin for similar conditions. During these time periods the downslope flows appears to be free from the interactions with the valley circulation.

4.1 Vertical Profiles of Turbulence Quantities

One of the six episodes of the katabatic winds mentioned above (YD 272 from 0200 to 0400 UTC) we consider in detail (other episodes will be also analysed in the coming sections). Figures 3 and 4 show vertical profiles of the half-hourly average wind speed, air temperature, turbulent fluxes, standard deviation of the sonic temperature, turbulent kinetic energy, and dissipation rate of turbulent kinetic energy observed at the ES3 flux tower on the East slope of Granite Mountain for four different time periods during the night of 28 September 2012 (YD 272 from 0200 to 0400 UTC); local time is from 2000 to 2200 of the previous day, 27 September 2012. In Fig. 5 we compare the average profiles of mean wind speed, the downwind stress (momentum flux), and standard deviation of air temperature measured at the ES4 tower with their counterparts observed at the ES3 tower for the same time periods (Figs. 3 and 4). Moreover, Fig. 5d shows vertical profiles of the dissipation (destruction) rate for half the temperature variance, $N_t$, observed at the ES4 tower. Unlike plots in Figs. 3-5, where the temporal evolution of turbulence vertical profiles at the ES3 and ES4 towers are presented (Eulerian description of the slope flow), Fig. 6 shows spatial behaviour of the vertical profiles of the half-hourly average wind speed and turbulent fluxes along the slope at the four ES2-ES5 flux towers for specific time period JD 272, 0300 UTC (2-D description). Turbulent fluxes and variances in Figs. 3-7 are computed through the frequency integration over the high-frequency parts of the appropriate spectra and cospectra (with about a 2-min cut-off time scale as the low-pass filter) that are associated with energy-containing/flux-carrying eddies (see Sect. 3). Turbulent quantities based on the high-frequency parts of the spectra and cospectra provide somewhat smaller scatter of the data as compared to their half-hourly average counterparts.

The vertical profiles of downslope wind speed from all sites (Figs. 3a, 5a, 6a) show a typical "pure" katabatic flow structure with wind-speed maximum located between 3 and 5 m. Figure 3b shows a typical vertical profile of air temperature measured by the 'slow' T/RH probes. Note slow cooling of the air layer for four different time periods during 0200-0330 UTC (Fig. 3b). During the time covered by Fig. 3, relative humidity at ES3 tower monotonically decreases from 30-36% at the 0.5 m measurement level to 23-25% in the layer 10-20 m (not shown).

According to Figs. 3-6, the profiles of velocity, turbulent fluxes, and other quantities show steep gradients in the layer below a wind-speed maximum. Obviously in this region the concept of the constant-flux layer is invalid for momentum and heat fluxes. However above the slope jet, the wind speed, temperature, turbulent fluxes, and variances vary with height more slowly than near the surface (approximately an order of magnitude). In the region of a wind-speed maximum, a local minimum is found for the TKE, $<e>=\left(\sigma_u^2+\sigma_v^2+\sigma_w^2\right)/2$, (Fig. 4c) and the dissipation rate of TKE (Fig. 4d), whereas the standard deviation of the sonic temperature, $\sigma_t$, has an absolute maximum near a wind-speed maximum (Figs. 4b and 5c). Although this behaviour of the TKE and $\sigma_t$ has been previously predicted by Horst and Doran (1988), Denby (1999, Figs. 3-4), and Söderberg and Parmhed (2006), a reliable experimental verification for katabatic flows was lacking. The dissipation (destruction) rate for half the temperature variance, $N_t$, derived from Eq. (7) generally decreases monotonically with height, although several cases of a weak local maximum near a wind-speed maximum were found (Fig. 5d).

In the case of a nocturnal low-level jet (LLJ), Banta et al. (2006) and Pichugina and Banta (2010) reported the minimum in $\sigma_u^2$ (and TKE) at the LLJ nose observed by the high resolution Doppler lidar. The minimum in TKE at the jet nose results from $<u'w'>\left(\partial U/\partial n\right)$ becoming 0 at this level, as noted by Banta et al. (2006, p. 2716). Although, the shear production of TKE is vanished at the level of the wind-speed maximum, but it is not necessarily true that TKE and $\varepsilon$ also vanishes there, or that no vertical mixing occurs through this level. Data in the present study and in Banta et al. (2006) indicate that TKE (or $\sigma_u^2$) and $\varepsilon$ became small but remained nonzero at the height of wind-speed maximum.

As mentioned earlier in Sect.2, a striking feature of katabatic flows is a sign reversal of the vertical momentum flux (downslope stress), $\tau = -<u'w'>$, and the horizontal temperature (heat) flux, $<u'\theta_v'>$, at the wind-maximum height. Observed profiles of $<u'w'>$ and $<u'\theta'>$ over the East slope of Granite Mountain are shown in Figs 3c, 5b, 6b and Figs. 4a, 6d respectively. According to our data, $<u'w'>$ is negative (positive) whereas $<u'\theta'>$ is positive (negative) below (above) the wind speed maximum. In other words, the vertical momentum flux is directed downward (upward) whereas the horizontal temperature flux is downslope (upslope) below (above) the wind-speed maximum in a slope-following coordinate system. Therefore, we suggest that the position of the jet-speed maximum can be derived from Figs. 3-6 using the intersection of linearly interpolated lines for positive and negative values of $<u'w'>$ or $<u'\theta_v'>$ with a vertical line (see next section for detail).

According to Figs. 3-5, the vertical profiles of the wind speed and various turbulent quantities are approximately stationary in time (especially near the surface) for specific tower (ES3 or ES4) during four different time periods for YD 272, 0200-0400 UTC. For example, vertical profiles of mean wind speed measured at the ES3 flux tower are almost identical for 0300 and 0330 (Fig.3a). However, according to Fig.6, the vertical profiles of the wind speed and turbulent fluxes along the flow line (from one tower to another) vary widely for fixed time period (JD 272, 0300 UTC). Note that significantly higher the momentum flux observed at the ES4 tower (Figs. 5b and 6b) which may be associated with higher aerodynamic roughness near the ES4 location (e.g., boulders, bushes). Thus, surface values of the turbulent fluxes in katabatic flows vary along a slope due to different properties of underlying surface. Remarkably, however, that the surface fluxes are almost constant over time for a specific slope location, implying that the katabatic flow adapts readily to new surface conditions down the slope.

Since the horizontal heat (buoyancy) flux $<u'\theta_v'>$ contributes to the net buoyancy term in the TKE budget equation (1) and that its observations still remain extremely limited, we consider its relative contribution to the buoyancy in more detail. Figure 7 shows the vertical profiles of the ratio $<u'\theta_v'>/<w'\theta_v'>$ measured at different towers during different days. According to Fig 7, the ratio $<u'\theta_v'>/<w'\theta_v'>$ has a negative minimum (positive maximum) below (above) the wind-speed maximum (Figs. 7b-d). Although typical values of the positive maximum for this ratio range between 5 and 10, some values can reach 13-19 (Figs. 7a-c). At the East slope of Granite Mountain values of $\cot\alpha$ range from 28.6 to 14.3 ($\alpha \approx$ 2-4°); this implies that the net buoyancy term $\beta(<w'\theta_v'>\cos\alpha - <u'\theta_v'>\sin\alpha)$ approximately equals to zero for $<u'\theta_v'>/<w'\theta_v'> \approx 19$ or even this term becomes positive, see the inequality (4). Thus, our data provides experimental evidence that the horizontal heat (buoyancy) flux in a slope-following coordinate system plays a crucial role in the dynamics of the katabatic winds even over gentle slopes.

4.2 Analysis of Turbulence Spectra and Cospectra

Figure 8 shows typical raw cospectra for the downwind stress (momentum flux) and the downwind horizontal flux of sonic temperature (kinematic horizontal sensible heat flux) at five levels (0.5, 2, 5, 10, and 20 m) for a case of westerly katabatic flow observed at the ES3 flux tower on 28 September 2012 (YD 272, 0230 UTC); local time is 2030 of the previous day. True wind direction derived from the sonic anemometers is in the range 273-286° for all five levels. Frequency-weighted cospectra in Fig. 8 are in log-linear coordinates, so that the area under the spectral curve represents the total covariance. For the data shown in Fig. 8, the momentum flux $<u'w'>$ (m$^2$ s$^{-2}$) equals $\approx$ –0.0044, –0.0024, 0.0033, 0.0095, 0.0094 for the levels from 1 to 5 (cf. Fig. 8a); the vertical flux of sonic temperature $<w'\theta'>$ (K m s$^{-1}$) equals $\approx$ –0.0089, –0.0058,

–0.0039, –0.0058, –0.0023 for the levels from 1 to 5 (cospectra of $<w'\theta'>$ are not shown); the downwind horizontal flux of sonic temperature $<u'\theta'>$ (K m s$^{-1}$) equals $\approx$ 0.0430, 0.0395, –0.0407, –0.0162, –0.0062 for the levels from 1 to 5 (cf. Fig. 3b). The above flux values were derived through the frequency integration of the appropriate cospectra (see Fig. 8) computed from 27.31-min data blocks.

According to Fig. 8 and the above flux values, the momentum flux and the horizontal sensible heat flux change their sign between 2 m (level 2) and 5 m (level 3). In particular, the vertical momentum flux is directed downward (upward) below (above) the wind-speed maximum (cf. Smeets et al. 2000, their Fig. 4). As discussed earlier, this is associated with the fact that the levels 1 and 2 are located below a wind-speed maximum whereas the levels 3-5 are located above a wind-speed maximum. This is consistent with the vertical profile of the mean wind speed (m s$^{-1}$) $\approx$ 1.64, 2.34, 2.68, 2.01, and 1.12 (the levels from 1 to 5 respectively) for the case shown in Fig. 8. Therefore, we suggest deriving a position of the wind-speed maximum from linear interpolation between positive and negative values of the momentum flux (or horizontal heat flux). A height of the wind-speed maximum $H_{max}$ corresponds to a level where the fluxes $<u'w'>$ and $<u'\theta'>$ become zero. Based on the above values of $<u'w'>$ and $<u'\theta'>$, linear interpolation of the momentum flux between levels 2 and 3 gives $H_{max} \approx 3.26$ m whereas linear interpolation of the horizontal heat flux leads to 3.48 m (mean value for both methods $H_{max} \approx 3.37$ m). It is clear that the flux-interpolation method gives more accurate estimates of $H_{max}$ than a method based on measurements of vertical profile of the mean wind speed. In our case, this method leads to $H_{max}$ located somewhere between 2 and 5 meters. Note that an interpolation method can be applied only to variables which change sign at a level of the wind-speed maximum.

Figure 9 shows typical one-dimensional, raw energy spectra of the longitudinal, lateral, and vertical velocity components, and the sonic temperature for a case of westerly katabatic flow observed at flux tower, levels 2-6 (2, 5, 10, 20, and 28 m; level 1 at 0.5 m is missing), 28 September 2012 (YD 272, 0330 UTC); local time is 2130 of the previous day. The vertical profile of the mean wind speed (m s$^{-1}$) for the case shown in Fig. 6 is ≈ 2.32, 2.98, 2.22, 1.24, and 0.76 (the levels from 2 to 6 respectively). True wind direction derived from the sonic anemometers is in the range 276-285° for all five levels.

The case shown in Fig. 9 is interesting because the measurement level 3 located at 5 m is close to a wind-speed maximum. For this case (Fig. 9), the momentum flux $<u'w'>$ (m$^2$ s$^{-2}$) equals ≈ –0.0337, 0.0016, 0.0237, 0.0158, 0.0130 for the levels from 2 to 6; the vertical flux of sonic temperature $<w'\theta'>$ (K m s$^{-1}$) equals ≈ –0.0278, –0.0117, –0.0132, –0.0059, –0.0037 for the levels from 2 to 6; the downwind horizontal flux of sonic temperature $<u'\theta'>$ (K m s$^{-1}$) equals ≈ 0.0684, –0.0098, –0.0491, –0.0209, –0.0119 for the levels from 2 to 6. Accordingly, the level 2 is located below a wind-speed maximum whereas the levels 4-6 are located above it. Turbulent fluxes $<w'u'>$ and $<u'\theta'>$ at the measurement level 3 are close zero and, thus, this level is located close to a wind-speed maximum (slightly above). A height of the wind-speed maximum $H_{max}$ based on the linear interpolation of the momentum flux between levels 2 and 3 gives $H_{max}$ ≈ 4.86 m whereas linear interpolation of the horizontal heat flux leads to 4.62 m (mean value for both methods $H_{max}$ ≈ 4.74 m).

According to Fig. 9, the turbulent spectral curves have a wide inertial subrange, which displays the –5/3 Kolmogorov power law for velocity components (the Obukhov-Corrsin law for the passive scalar) at high frequencies (a slope of –2/3 for the frequency-weighted spectra plotted in Fig. 6) at all five sonic levels 2-6. Although in the layer above the flow wind maximum (levels 3-6), the momentum flux is negative, $\tau = -<u'w'> < 0$ (upward momentum transfer) and moreover at level 3 the production of TKE $-<u'w'>(\partial U/\partial n) \approx 0$, the turbulence here is still associated with the Richardson-Kolmogorov cascade (Fig. 9). Additional plots of the spectra and cospectra for katabatic winds can be found in Smeets et al (2000, Fig. 4). For the spectra shown in Fig. 9, the standard deviation of the longitudinal wind speed component $\sigma_u$ (m s$^{-1}$) equals $\approx$ 0.4144, 0.3538, 0.4391, 0.3614, and 0.3550 (levels from 2 to 6 respectively), the standard deviation of the lateral wind speed component $\sigma_v$ (m s$^{-1}$) equals $\approx$ 0.2944, 0.1983, 0.2952, 0.2887, and 0.2459 (levels from 2 to 6 respectively), the standard deviation of the vertical wind speed component $\sigma_w$ (m s$^{-1}$) equals $\approx$ 0.2137, 0.1650, 0.2048, 0.1926, and 0.1571 (levels from 2 to 6 respectively), and the standard deviation of the sonic temperature $\sigma_t$ (K) equals $\approx$ 0.3291, 0.5592, 0.2069, 0.1157, and 0.0794 (levels from 2 to 6 respectively). The above turbulent covariance and variance values were derived through the frequency integration of appropriate cospectra and spectra (Fig. 9) computed from 27.31-min data blocks (corresponding to $2^{15}$ data points).

According to the spectra plots in Fig. 9 and the above data, the standard deviations of all wind speed component (and therefore TKE) have a local minimum (cf. Fig. 4c), whereas the standard deviation of the air temperature $\sigma_t$ has an absolute maximum at the height of the slope wind maximum $H_{max}$ (cf. Figs. 4b and 5c). Figures 8 and 9 also supports the conclusion that the turbulent fluxes and variances in the layer below the wind-speed maximum vary with height faster (approximately an order of magnitude) than in the layer above the slope jet.

4.3 Local z-less Stratification

In the region of the wind-speed maximum, production of turbulence by wind shear is quite small or even zero at this maximum where $<u'w'> = 0$ and a local minimum in TKE and $\varepsilon$ (Figs. 4c, d) is observed. This suggests that turbulent exchange across the wind-speed maximum would cease and the turbulence above the slope jet can be largely decoupled from the flow below and underlying surface (Horst and Doran 1988; Denby 1999). Thus, in this region, the turbulence no longer communicates effectively with the surface and various quantities become independent of height of measurement $z$ (or $n$), that is, $z$ (or $n$) ceases to be a scaling parameter. This limit is termed 'z-less stratification' (height-independent) by Wyngaard and Coté (1972). Note that the difference between $z$ and $n = z\cos\alpha$ in our case is negligible (less than 1%). Thus $n \approx z$ and $\zeta = n/L \approx z/L$ for current study.

We tested the classical local z-less predictions for the Monin-Obukhov non-dimensional functions $\varphi_m$, $\varphi_\varepsilon$, and $\varphi_\alpha$ in the layer above the slope jet. The non-dimensional vertical gradient of the mean wind speed, $U$, and the non-dimensional dissipation rate of turbulent kinetic energy $\varepsilon$ according to Monin–Obukhov similarity theory (MOST) are expressed as:

$$\varphi_m(\zeta) = -\left(\frac{\kappa n}{u_*}\right)\frac{dU}{dn} \quad, \tag{8}$$

$$\varphi_\varepsilon(\zeta) = \frac{\kappa n \varepsilon}{u_*^3} \quad. \tag{9}$$

Note that the function $\varphi_m$ is defined $> 0$ above a wind-speed maximum. The standard deviations of wind speed components $\sigma_\alpha$ are scaled as

$$\varphi_\alpha(\zeta) = \frac{\sigma_\alpha}{u_*}, \qquad (10)$$

where $\alpha$ (= $u$, $v$, and $w$) denotes the longitudinal, lateral, or vertical velocity component, the friction velocity is computed as $u_* = \left(<u'w'>^2 + <v'w'>^2\right)^{1/4}$, and $\zeta = n/L$ is defined by (3). The $z$-less concept requires that $n$ cancels in Eqs. 8-10, which corresponds to

$$\varphi_m(\zeta) = \beta_m \zeta, \qquad (11)$$

$$\varphi_\varepsilon(\zeta) = \beta_\varepsilon \zeta, \qquad (12)$$

$$\varphi_\alpha(\zeta) = \beta_\alpha, \qquad (13)$$

where $\beta_m$, $\beta_\varepsilon$, and $\beta_\alpha$ are numerical coefficients. A simple linear interpolation $\varphi_m(\zeta) = 1 + \beta_m \zeta$ and $\varphi_\varepsilon(\zeta) = 1 + \beta_\varepsilon \zeta$ have been suggested to provide blending between neutral and very stable ('$z$-less') cases.

Figure 10 shows plots for the normalized standard deviations of all three wind speed components defined by Eq. 10 (local scaling). According to Fig. 10, the universal functions $\varphi_\alpha(\zeta)$ are approximately constant, that is, they are consistent with the classical Monin-Obukhov $z$-less prediction (13). The horizontal dashed lines in Fig. 10 correspond to $\beta_u = 2.3$, $\beta_v = 2.0$, and $\beta_w = 1.5$ which are median values computed for individual 1/2-hour averaged points. Note, that our plots in Fig. 10 are consistent with the results by Horst and Doran (1988, their Figs. 2, 3) and Smeets et al. (2000, their Fig. 10). Although the data presented in Fig. 10 generally proves the validity of the $z$-less approach (13), the plots in Fig. 10 are affected by self-correlation because $u_*$ appears both in the definitions of the universal functions $\varphi_\alpha$ and in $\zeta$. This results in a weak trend of the data points in Fig. 10. However, this flaw can be overcome by plotting $\varphi_\alpha$ versus a stability parameter which contains no $u_*$ (see Grachev et al. 2013, 2014 for discussion).

Figure 11 shows plots of the non-dimensional universal functions $\varphi_m$, Eq. 8 and $\varphi_\varepsilon$, Eq. 9, versus the Monin-Obukhov stability parameter for local scaling $\zeta = n/L \approx z/L$, Eq. 3. According to Fig. 11 our data are consistent with the linear interpolations $\varphi_m(\zeta) = 1 + \beta_m \zeta$ and $\varphi_\varepsilon(\zeta) = 1 + \beta_\varepsilon \zeta$ with numerical coefficients $\beta_m = 4.1$ (Fig. 11a) and $\beta_\varepsilon = 5.2$ (Fig. 11b); and, therefore, the data are consistent with the z-less predictions (11) and (12).

Similarly to plots in Figs. 3-7, turbulent fluxes and variances in Figs. 10 and 11 are computed through the frequency integration over the high-frequency parts of the appropriate spectra and cospectra. Because here we consider only a region above the slope jet, data collected at levels 3-7 of the ES2 tower, levels 3-5 of the ES3 and ES5 towers, and levels 4-6 of the ES4 tower are only analysed in Figs. 10 and 11. All six cases of westerly katabatic winds mentioned in Sec. 4 are used in Figs. 10 and 11 (records were only accepted if the true wind direction at all towers and all levels was within a 280±30° sector).

Furthermore, the data presented in Figs. 10 and 11 were quality controlled prior to evaluating similarity functions (8)-(10) in order to remove spurious or low-quality records. The following filtering criteria are adopted in this study (see Grachev et al. 2013, 2014) and references therein for discussion). To avoid a possible flux loss caused by inadequate frequency response and sensor separations, we omitted data with a local wind speed less than 0.2 m s$^{-1}$. We set minimum or/and maximum thresholds for the kinematic momentum flux (> 0.0002 m$^2$ s$^{-2}$), vertical and horizontal temperature fluxes (<–0.0002 K m s$^{-1}$), standard deviation of each wind speed component (>0.01 m s$^{-1}$), standard deviation of air temperature (>0.01 K), vertical gradients of mean velocity (<–0.001 s$^{-1}$), dissipation rate of turbulent kinetic energy (0.00002 < $\varepsilon$ < 0.1 m$^2$ s$^{-3}$) and the dissipation (destruction) rate for half the temperature variance (0.00002 <

$N_t < 0.01$ K$^2$ s$^{-1}$). Points with excessive standard deviation of wind direction (>30°), steadiness (trend) of the non-rotated wind speed components ($\Delta u / U < 1$, $\Delta v / U < 1$), and sonic temperature (>2°C) were also removed to avoid non-stationary conditions during a 1/2-hour record. In addition, sonic anemometer angle of attack was limited by 10°.

**5 Summary and Conclusions**

In this study, we discussed the small-scale turbulence structure of katabatic flows based on tower measurements made over the complex-terrain during the first MATERHORN field campaign (MATERHORN–Fall) at the US Army Dugway Proving Grounds in Utah from 25 September through 31 October, 2012. The MATERHORN field campaign is the most comprehensive study thus far on the small-scale katabatic flows. Turbulent and mean meteorological data were collected at multiple levels (up to seven) on four ES2-ES5 flux towers deployed along East slope (2-4°) of Granite Mountain (Figs. 1 and 2). This allows studying temporal and spatial structure of nocturnal slope flows in detail providing insights into the nature of the phenomenon.

Katabatic winds develop soon after sunset when the surface starts to cool, and they are associated with quiescent synoptic conditions and clear skies. In general, these winds are considered to be unidirectional and persistent. It is found, however, that westerly katabatic flows over the East slope of Granite Mountain are rather intermittent due to interactions with valley flows occurring at various times during the night. The flow appears to be free from those interactions generally only soon after sunset, over a duration of about 2-3 hours. In this study we analysed only such flows that resemble a "pure" katabatic flow structure at all ES2-ES4 flux towers at the same time.

The most prominent feature of katabatic flows is a wind-speed maximum close to the surface (Figs. 3a, 5a, 6a) that causes change in sign for the vertical momentum flux (downslope stress), $\tau = -<u'w'>$, and the horizontal heat (buoyancy) flux, $<u'\theta'_v>$ below and above this maximum. According to our data, $<u'w'>$ is negative (positive) whereas $<u'\theta'>$ is positive (negative) below (above) the wind speed maximum (Figs 3c, 5b, 6b and Figs. 4a, 6d respectively). In other words, the vertical momentum flux is downward (upward) whereas the horizontal temperature flux is downslope (upslope) below (above) the wind-speed maximum in a slope-following coordinate system. Our study suggests that the position of jet speed maximum can be derived from linear interpolation between positive and negative values of the momentum flux (or the horizontal heat flux) and determination of the height where flux becomes zero. Furthermore, it is shown that the standard deviations of all wind speed component (and therefore TKE) and the dissipation rate of TKE have a local minimum (Figs. 4c, d), whereas the standard deviation of air temperature $\sigma_t$ has an absolute maximum (Figs. 4b and 5c) near a wind-speed maximum.

It is found that the profiles of velocity, turbulent fluxes, and other quantities have steep gradients in the layer below a wind-speed maximum (Figs. 3-6). Above the slope jet, however, the wind speed, temperature, turbulent fluxes, and variances vary with height more slowly than near the surface (approximately an order of magnitude). According to our data (Figs. 3-5), the vertical profiles of wind speed, turbulent fluxes, and variances are approximately stationary in time (especially near the surface) for a given tower during specific time intervals. However, the vertical profiles of wind speed and turbulent fluxes along the tower line vary widely for given time period, characterizing the spatial evolution (Fig. 6).

The slope flows are traditionally described and modelled in a slope-following coordinate system. Because the mean flow is not normal to the direction of gravity, the buoyancy term in turbulence equations include extra terms associated with the horizontal (slope-parallel) heat flux, $\beta <u'\theta_v'> \sin\alpha$, which can enhance or suppress turbulence. The horizontal heat flux is a sink (source) for TKE below (above) the wind maximum, and, therefore, in slope flows $Rf$ and $z/L$ below (above) the wind maximum is smaller (larger) than over flat horizontal surfaces. Moreover, we describe several cases when $<u'\theta_v'>/<w'\theta_v'> \approx 19$ (Figs. 7a, b), implying that the net buoyancy term $\beta(<w'\theta_v'>\cos\alpha - <u'\theta_v'>\sin\alpha) \approx 0$ for typical slopes at the East slope site ($\alpha \approx 2\text{-}4°$). In this case the destructive effect of vertical heat (buoyancy) flux is completely cancelled by the generation of turbulence due to the horizontal heat (buoyancy) flux.

The zero wind shear, change in the sign of vertical momentum flux, local minimum in TKE and dissipation rate, and the background stable stratification suggest that turbulence in the layer above the wind-speed maximum is decoupled from the surface. In other words, turbulence no longer communicates significantly with the surface, making the height of the measurement $z$ irrelevant as a governing parameter. We hypothesize that turbulence in this layer is consistent with the classical local $z$-less (height-independent) predictions for stably stratified boundary layer. The normalized standard deviations of all three wind speed components, the non-dimensional vertical gradient of mean wind speed, and the non-dimensional dissipation rate of turbulent kinetic energy were in good agreement with the $z$-less concept (Figs. 10 and 11).

**Acknowledgements**   The MATERHORN Program was funded by the Office of Naval Research with award # N00014-11-1-0709, with additional funding from the Army Research Office, Air Force Weather Agency, University of Notre Dame, University of Utah. Special thanks go to Evgeni Fedorovich who pointed out the importance of the horizontal heat (buoyancy) flux, $\beta <u'\theta_v'> \sin\alpha$, in the net buoyancy term in the TKE equation (1) and in the modified Monin-Obukhov stability parameter (3).

**Figure Captions**

Fig. 1. Aerial view of the topography of Granite Mountain showing a schematic view of the experimental set-up at the East slope (towers ES2-ES5).

Fig. 2. Elevation cross-section at the location of the ES2-ES5 flux towers on the East slope of Granite Mountain.

Fig. 3. Plots of vertical profiles of the (*a*) wind speed, (*b*) air temperature (level 2 is missing), (*c*) $<u'w'>$, (*d*) $<w'\theta'_v>$ observed at the ES3 flux tower on the East slope of Granite Mountain on 28 September 2012 (YD 272, 0200-0330 UTC).

Fig. 4. Plots of vertical profiles of the (*a*) $<u'\theta'_v>$ (*b*) standard deviation of the air temperature, $\sigma_t$, (*c*) turbulent kinetic energy (TKE), (*d*) dissipation rate of TKE, $\varepsilon$, observed at the ES3 flux tower on 28 September 2012 (YD 272, 0200-0330 UTC).

Fig. 5. Plots of vertical profiles of the (*a*) wind speed, (*b*) $<u'w'>$, (*c*) standard deviation of the air temperature, $\sigma_t$, (*d*) dissipation (destruction) rate for half the temperature variance observed at the ES4 flux tower (level 1 is missing) on 28 September 2012 (YD 272, 0200-0330 UTC).

Fig. 6. Plots of vertical profiles of the (*a*) wind speed, (*b*) $<u'w'>$, (*c*) $<w'\theta'_v>$, (*d*) $<u'\theta'_v>$ observed at the ES2, ES3, ES4, and ES5 flux towers on the East slope of Granite Mountain on 28 September 2012 (YD 272, 0300 UTC).

Fig. 7. Plots of vertical profiles of the ratio $<u'\theta_v'>/<w'\theta_v'>$ for (*a*) ES2 tower, YD 274, 0200-0330 UTC, (*b*) ES3 tower, YD 272, 0200-0330 UTC, (*c*) ES3 tower, YD 276, 0200-0330 UTC, (*d*) ES5 tower, YD 272, 0200-0330 UTC.

Fig. 8. Typical raw cospectra of (*a*) the downwind stress and (*b*) the downwind horizontal flux of sonic temperature at five levels (0.5, 2, 5, 10, and 20 m) for a case of katabatic flow observed at the ES3 flux tower on the East slope of Granite Mountain on 28 September 2012 (YD 272, 0230 UTC); local time is 2030 of the previous day (local time zone is MDT). The cospectra are computed from 27.31-min data blocks (corresponding to $2^{15}$ data points). Note that the levels 1 and 2 are located below the wind-speed maximum whereas the levels 3-5 are located above the wind-speed maximum.

Fig. 9. Typical raw energy spectra of the (*a*) longitudinal, (*b*) lateral, and (*c*) vertical velocity components and (*d*) the sonic temperature for a case of katabatic flow observed at the ES4 flux tower, levels 2-6 (2, 5, 10, 20, and 28 m; level 1 at 0.5 m is missing), 28 September 2012 (YD 272, 0330 UTC); local time is 2130 of the previous day (local time zone is MDT). The spectra are computed from 27.31-min data blocks (corresponding to $2^{15}$ data points). Note that the level 2 is located below the wind-speed maximum whereas the levels 4-6 are located above a wind-speed maximum. The measurement level 3 is close to the wind-speed maximum.

Fig. 10. The non-dimensional standard deviations of (*a*) longitudinal (down-slope), (*b*) lateral (cross-slope), and (*c*) vertical (normal) wind speed component (local scaling) observed for katabatic winds in the layer above the slope jet at the ES2-ES5 flux towers on the East slope of Granite Mountain. The horizontal dashed lines correspond to $\beta_u = 2.3$, $\beta_v = 2.0$, and $\beta_w = 1.5$.

Fig. 11. Same as Fig. 10 but for the non-dimensional universal functions (*a*) $\varphi_m$ and (*b*) $\varphi_\varepsilon$. The dashed lines are based on $\beta_m = 4.1$ and $\beta_\varepsilon = 5.1$. Note that the function $\varphi_m$ is defined as positive for negative vertical gradients of mean wind speed in the layer above a wind-speed maximum.

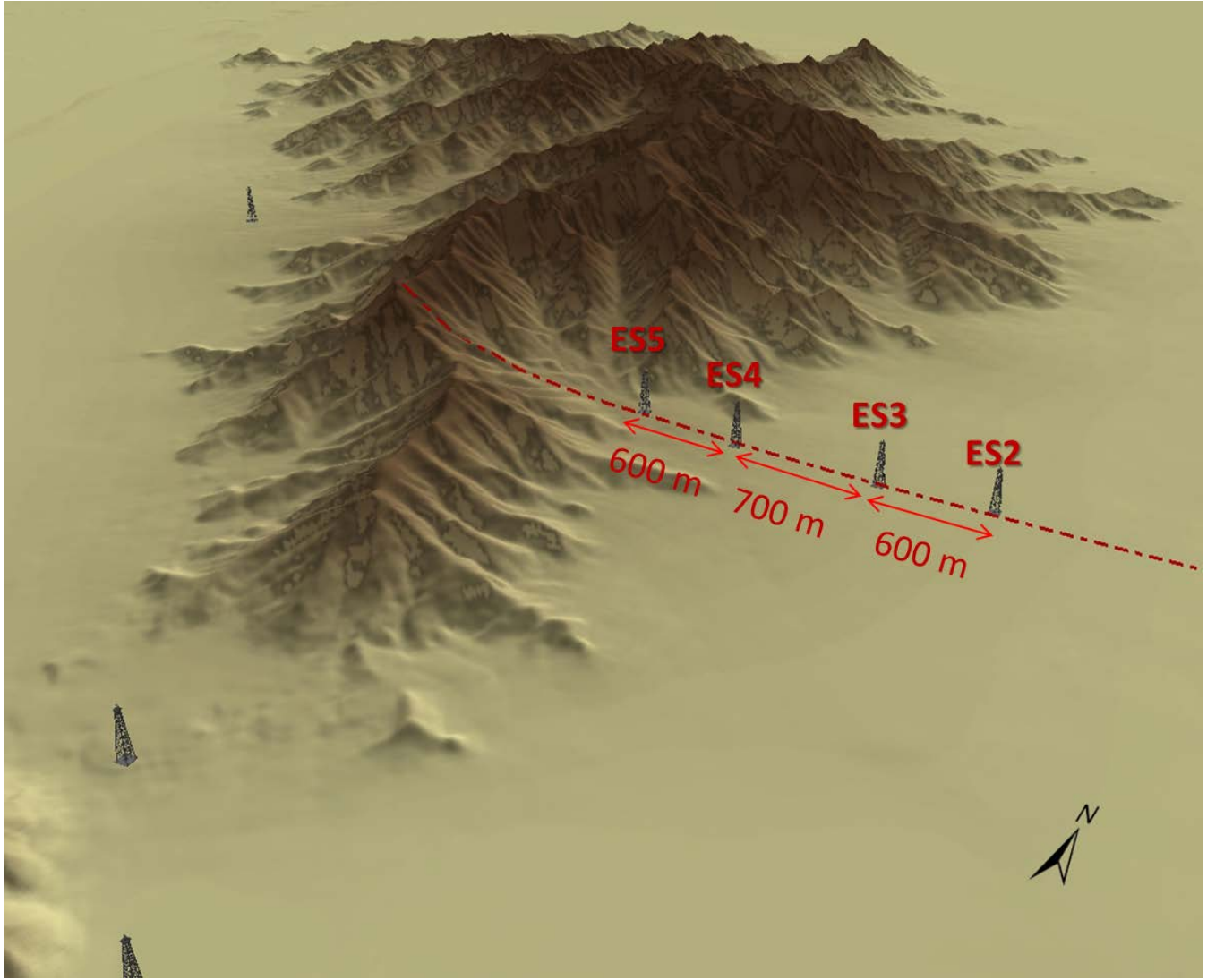

Fig. 1. Aerial view of the topography of Granite Mountain showing a schematic view of the experimental set-up at the East slope (towers ES2-ES5).

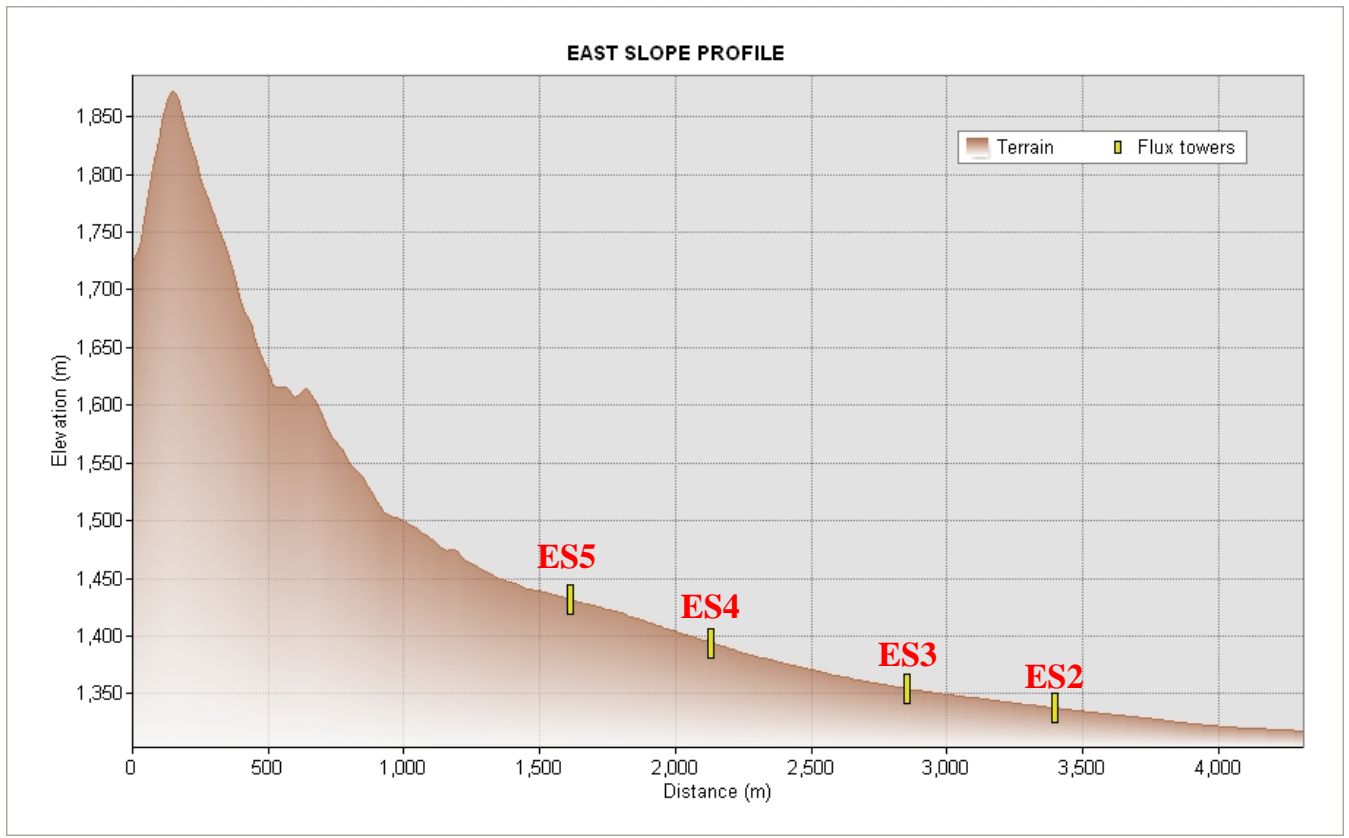

Fig. 2. Elevation cross-section at the location of the ES2-ES5 flux towers on the East slope of Granite Mountain.

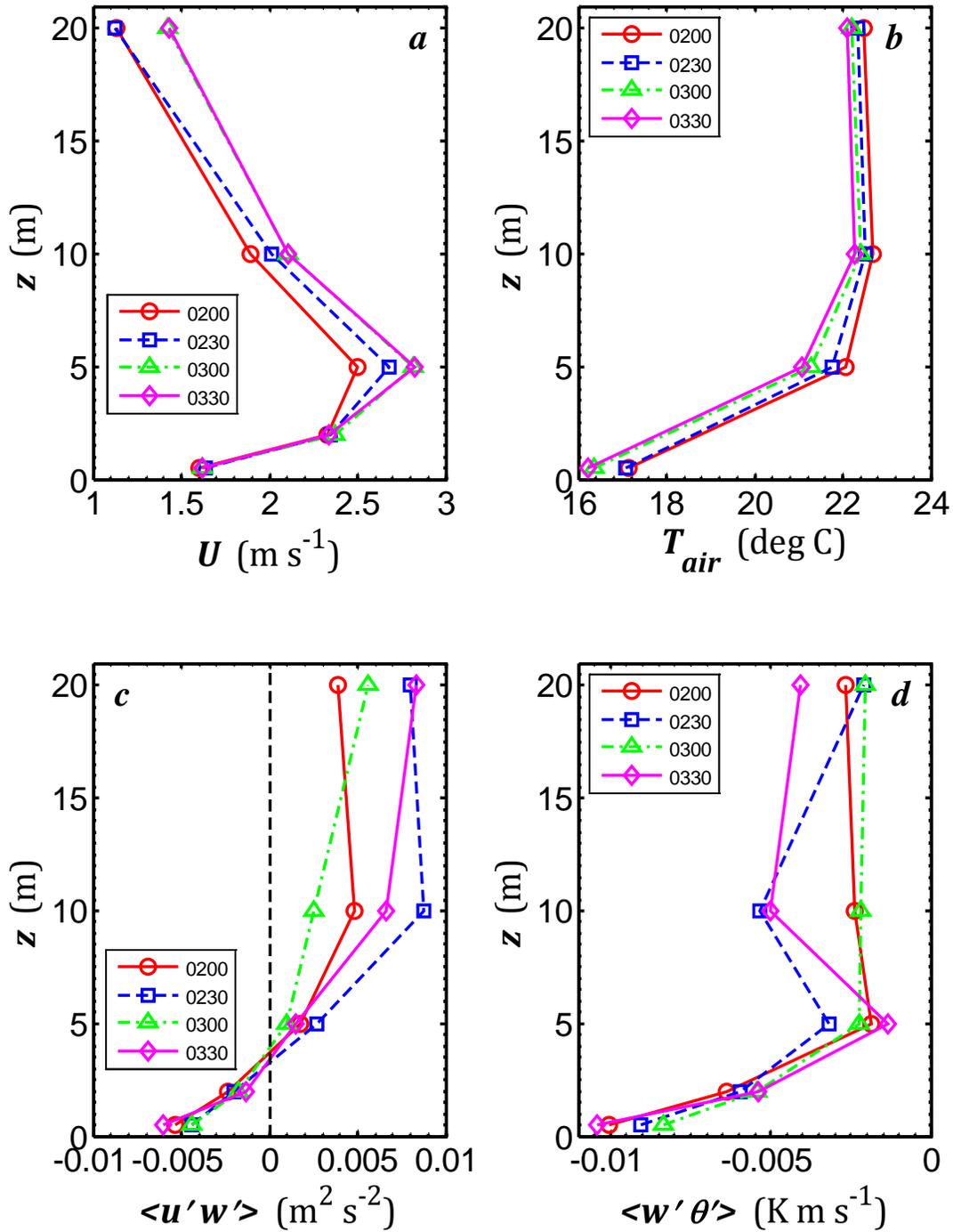

Fig. 3. Plots of vertical profiles of the (*a*) wind speed, (*b*) air temperature (level 2 is missing), (*c*) $<u'w'>$, (*d*) $<w'\theta'_v>$ observed at the ES3 flux tower on the East slope of Granite Mountain on 28 September 2012 (YD 272, 0200-0330 UTC).

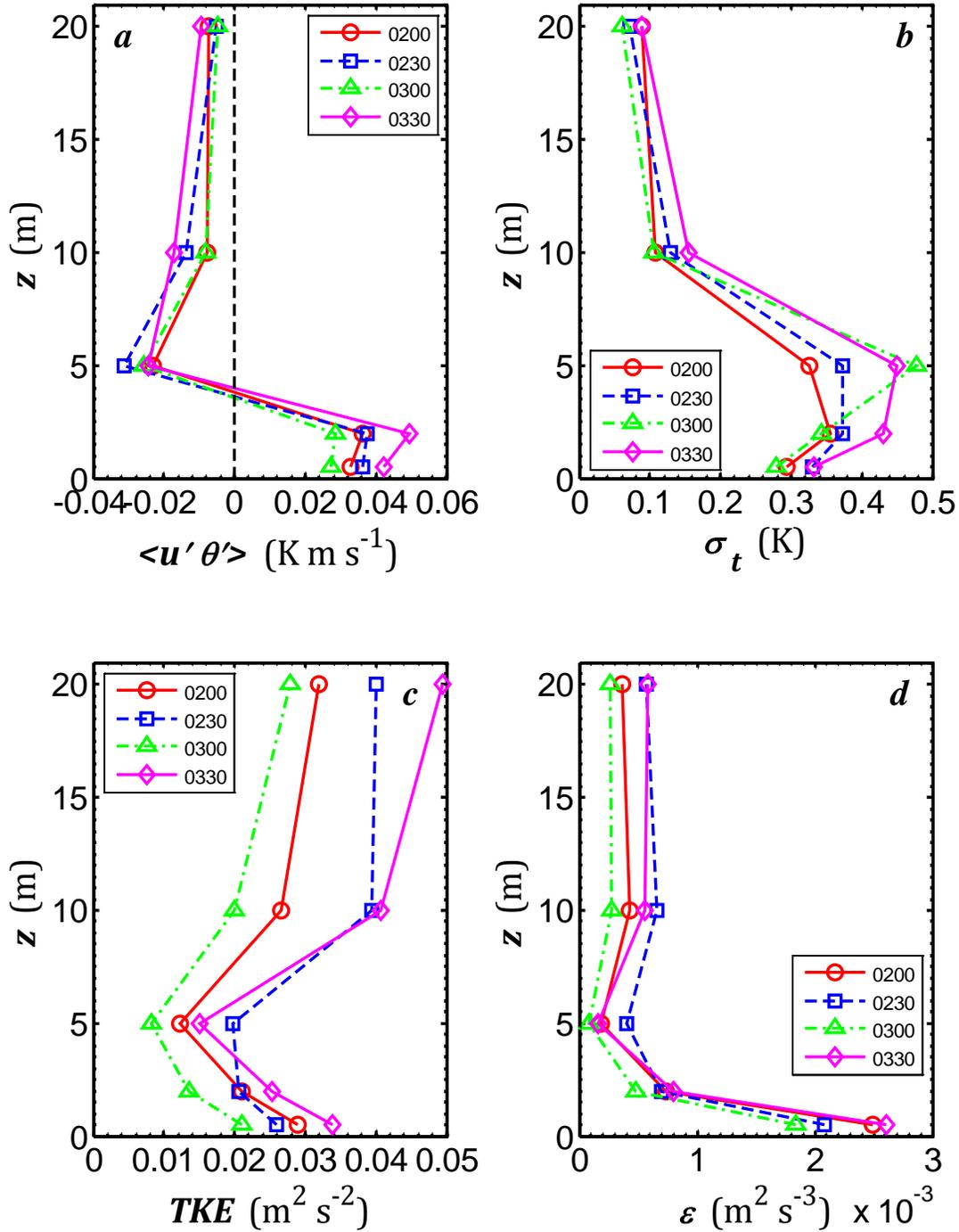

Fig. 4. Plots of vertical profiles of the (a) $<u'\theta'_v>$ (b) standard deviation of the air temperature, $\sigma_t$, (c) turbulent kinetic energy (TKE), (d) dissipation rate of TKE, $\varepsilon$, observed at the ES3 flux tower on 28 September 2012 (YD 272, 0200-0330 UTC).

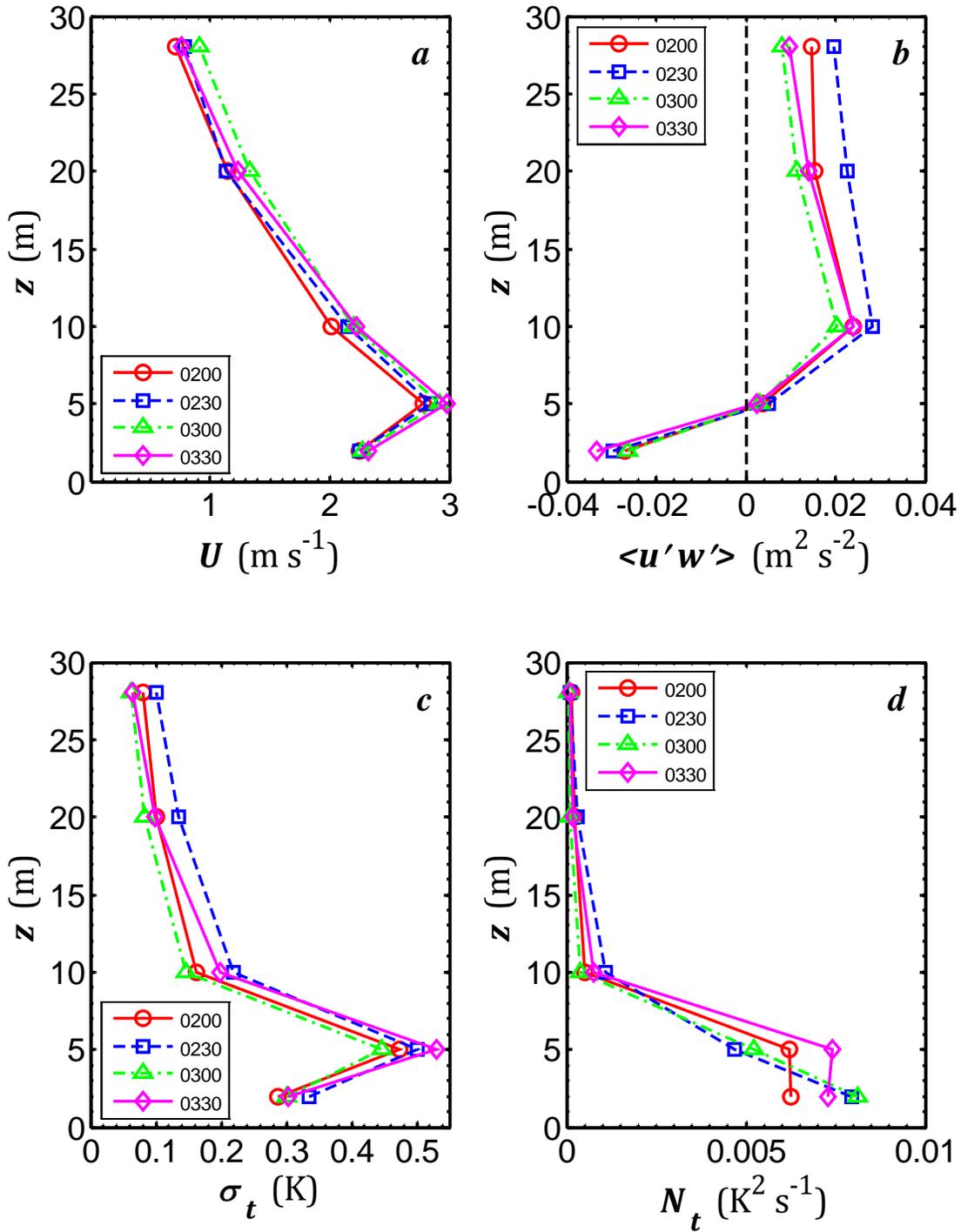

Fig. 5. Plots of vertical profiles of the (*a*) wind speed, (*b*) $<u'w'>$, (*c*) standard deviation of the air temperature, $\sigma_t$, (*d*) dissipation (destruction) rate for half the temperature variance observed at the ES4 flux tower (level 1 is missing) on 28 September 2012 (YD 272, 0200-0330 UTC).

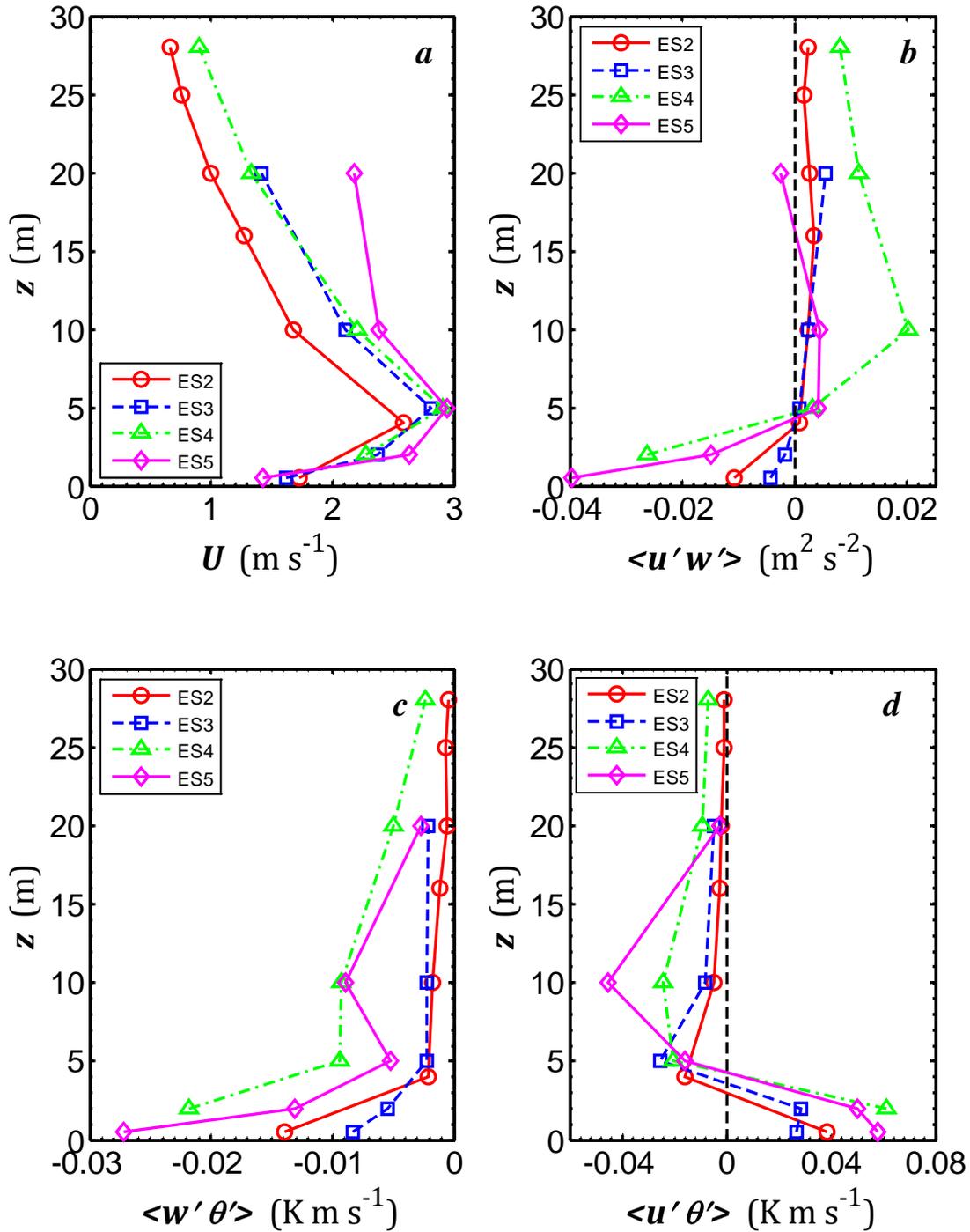

Fig. 6. Plots of vertical profiles of the (a) wind speed, (b) $<u'w'>$, (c) $<w'\theta_v'>$, (d) $<u'\theta_v'>$ observed at the ES2, ES3, ES4, and ES5 flux towers on the East slope of Granite Mountain on 28 September 2012 (YD 272, 0300 UTC).

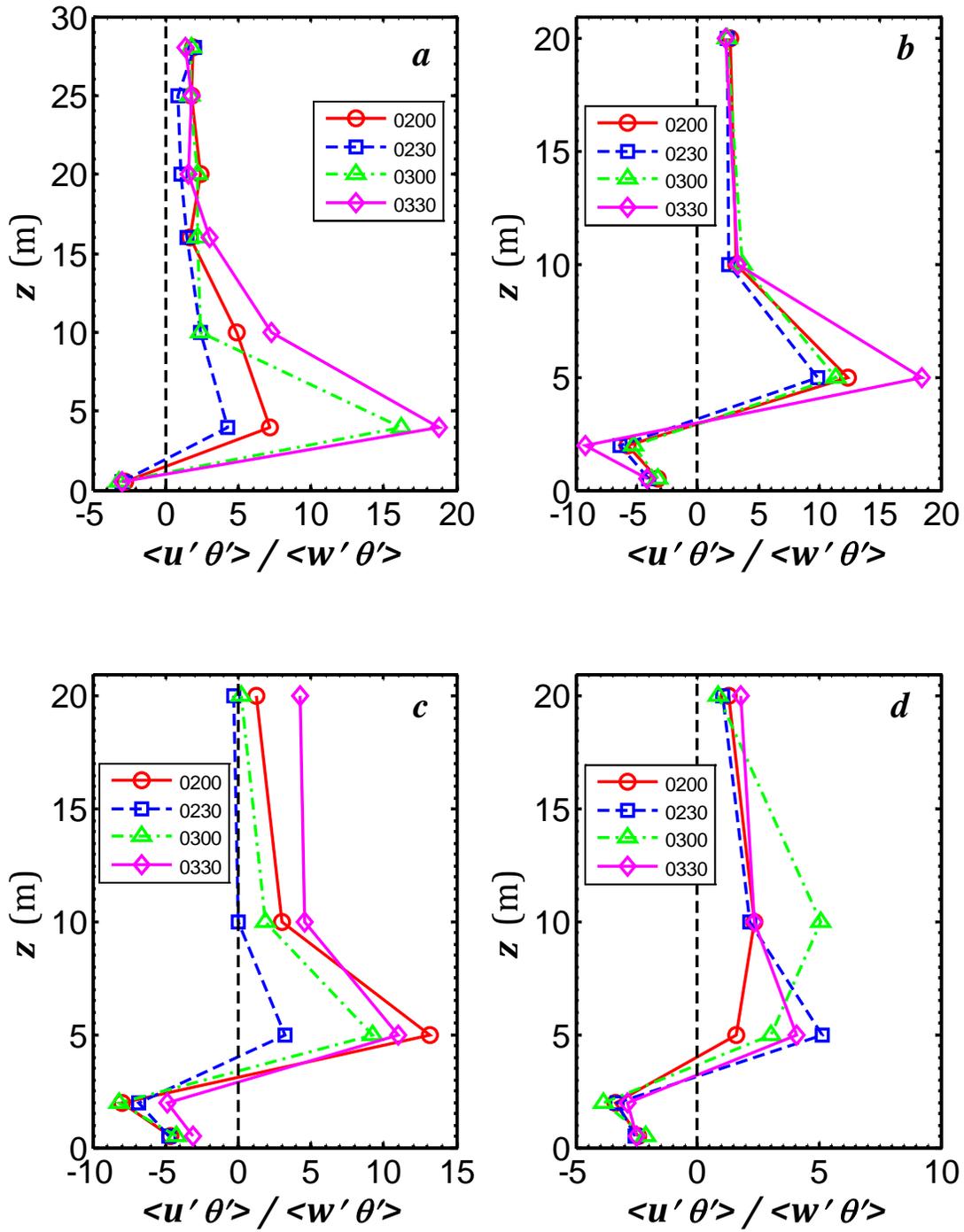

Fig. 7. Plots of vertical profiles of the ratio $<u'\theta'_v>/<w'\theta'_v>$ for (*a*) ES2 tower, YD 274, 0200-0330 UTC, (*b*) ES3 tower, YD 272, 0200-0330 UTC, (*c*) ES3 tower, YD 276, 0200-0330 UTC, (*d*) ES5 tower, YD 272, 0200-0330 UTC.

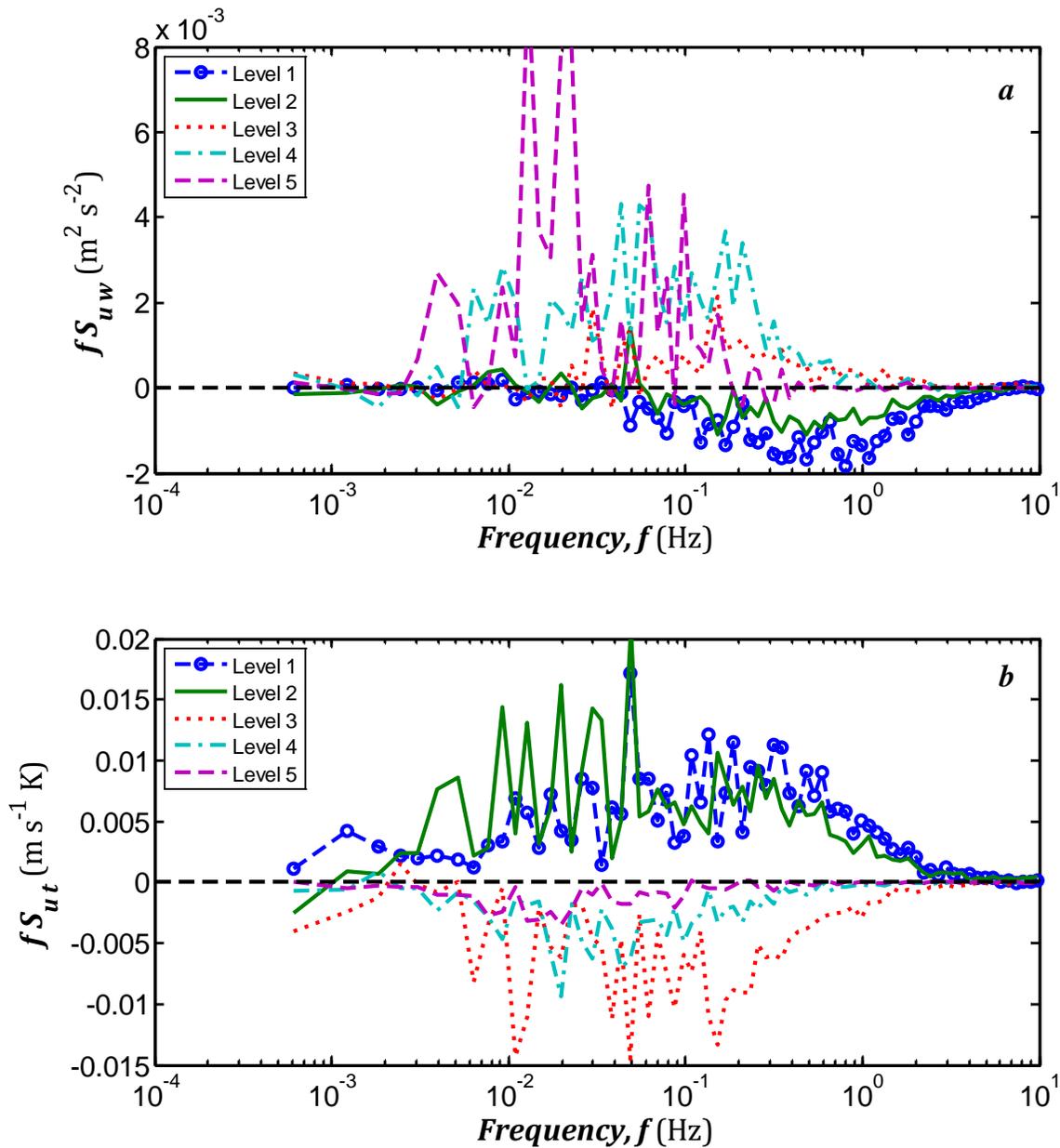

Fig. 8. Typical raw cospectra of (*a*) the downwind stress and (*b*) the downwind horizontal flux of sonic temperature at five levels (0.5, 2, 5, 10, and 20 m) for a case of katabatic flow observed at the ES3 flux tower on the East slope of Granite Mountain on 28 September 2012 (YD 272, 0230 UTC); local time is 2030 of the previous day (local time zone is MDT). The cospectra are computed from 27.31-min data blocks (corresponding to $2^{15}$ data points). Note that the levels 1 and 2 are located below the wind-speed maximum whereas the levels 3-5 are located above the wind-speed maximum.

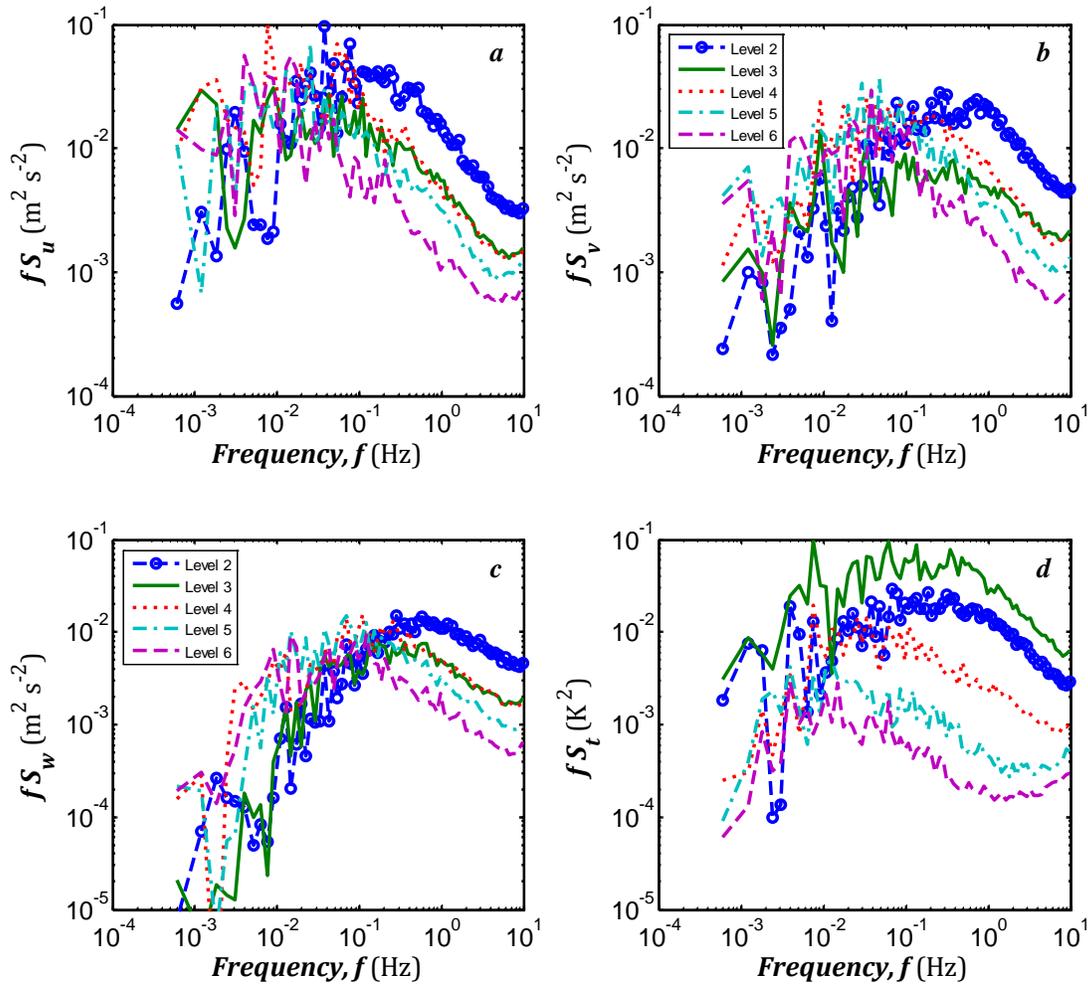

Fig. 9. Typical raw energy spectra of the (*a*) longitudinal, (*b*) lateral, and (*c*) vertical velocity components and (*d*) the sonic temperature for a case of katabatic flow observed at the ES4 flux tower, levels 2-6 (2, 5, 10, 20, and 28 m; level 1 at 0.5 m is missing), 28 September 2012 (YD 272, 0330 UTC); local time is 2130 of the previous day (local time zone is MDT). The spectra are computed from 27.31-min data blocks (corresponding to $2^{15}$ data points). Note that the level 2 is located below the wind-speed maximum whereas the levels 4-6 are located above a wind-speed maximum. The measurement level 3 is close to the wind-speed maximum.

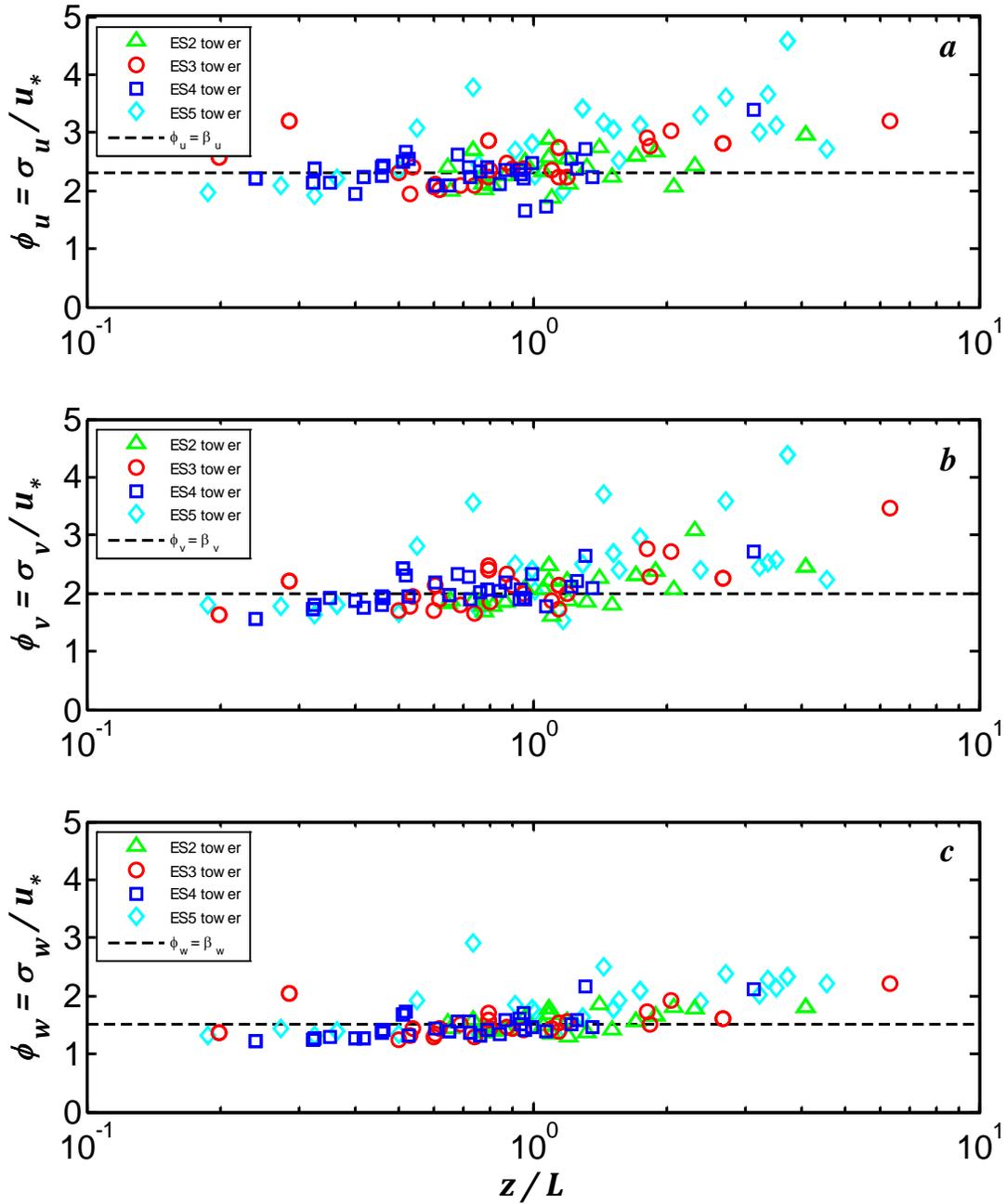

Fig. 10. The non-dimensional standard deviations of (*a*) longitudinal (down-slope), (*b*) lateral (cross-slope), and (*c*) vertical (normal) wind speed component (local scaling) observed for katabatic winds in the layer above the slope jet at the ES2-ES5 flux towers on the East slope of Granite Mountain. The horizontal dashed lines correspond to $\beta_u = 2.3$, $\beta_v = 2.0$, and $\beta_w = 1.5$.

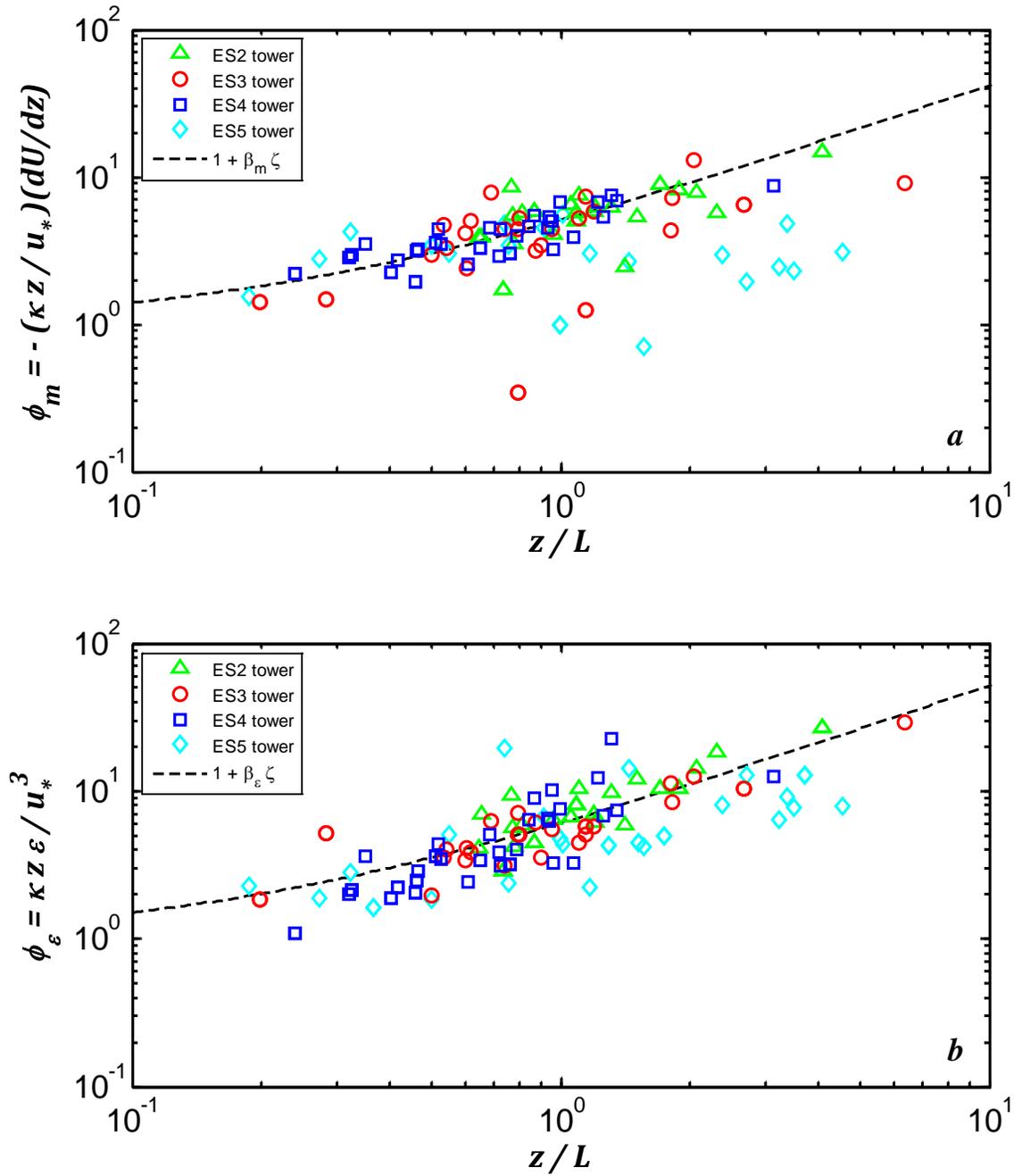

Fig. 11. Same as Fig. 10 but for the non-dimensional universal functions (*a*) $\varphi_m$ and (*b*) $\varphi_\varepsilon$. The dashed lines are based on $\beta_m = 4.1$ and $\beta_\varepsilon = 5.1$. Note that the function $\varphi_m$ is defined as positive for negative vertical gradients of mean wind speed in the layer above a wind-speed maximum.